\documentstyle[epsfig]{aipproc}

\def\CXO{{\it Chandra X-ray Observatory}}
\def\chandra{{\it Chandra}}
\def\xmm{{\it XMM-Newton}}
\def\EO{{\it Einstein Observatory}}
\def\einstein{{\it Einstein}}
\def\rosat{{\it ROSAT}}
\def\hubble{{\it Hubble Space Telescope}}
\def\hst{{\it HST}}
\def\asca{{\it ASCA}}

\begin{document}
\title{Young Supernova Remnants in the Magellanic Clouds}

\author{John P.~Hughes$^1$}
\address{$^1$ Rutgers University, Department of Physics and Astronomy}

\maketitle

\begin{abstract}
There are a half-dozen or so young supernova remnants in the Magellanic
Clouds that display one or more of the following characteristics: high
velocity ($\gtrsim$1000 km s$^{-1}$) emission, enhanced metallicity,
or a rapidly rotating pulsar. I summarize the current state of
knowledge of these remnants and present some recent results mostly
from the new X-ray astronomy satellites. 
\end{abstract}

\section*{Introduction}

The Magellanic Clouds (MCs) have been fertile hunting ground for
supernova remnants (SNRs) for the last 40 years or so.  Early work,
focused on the optical \cite{henize} and radio \cite{mathewsonhealy64}
bands, resulted in the discovery of the first SNRs in the MCs, N132D,
N49, and N63A. Since then deeper radio surveys as well as new X-ray
surveys \cite{lhg81} in conjunction with optical spectroscopic
follow-up have yielded a grand total of more than 50 confirmed
remnants in the Clouds.

The common, well-known distances to the Clouds (50 kpc and 60 kpc for
the large and small Clouds, respectively) allow for accurate estimates
of such quantities as mass and energy, which have strong dependences
on distance ($D^{5/2}$). The Clouds are close enough that angular
resolution is not a major limiting factor, particularly now with the
superb telescopes on the \hubble\ and \CXO.  The generally low
absorption to the Clouds makes possible UV and soft X-ray studies that
are often precluded for Galactic remnants and indeed nearly all MC
SNRs have been detected in the radio, optical, and X-ray bands.  The
sample of MC remnants is thus well-defined with reasonably clear
selection criteria.  With it one can sample the range of remnant types
and SN progenitors, as well as probe the dynamical evolution of SNRs.

Since SNRs can exist as distinct entities for tens to hundreds of
thousands of years, most of the remnants in the MCs are middle-aged or
old. But chronological age alone is not the only determinant of youth;
environment plays an equally important role.  In a low density
environment a remnant will expand to great size and the ejecta can
dominate the dynamics and the emission properties for a considerable
time.  On the other hand, remnant evolution will proceed rapidly in
the densest environments, passing quickly to the radiative phase.
Thus a definition of what constitutes a young SNR is needed.  Here I
will use the following, largely observational, definition.  A young
remnant is one with a rapidly rotating, high spin-down-rate pulsar;
high velocity oxygen-rich optical emission; other evidence for
supernova (SN) ejecta; or a kinematic age of roughly 2000 yr or less.
%
%
Note that I will not discuss SN1987A since it is the subject of an
entire article elsewhere in this volume \cite{mccray01}.

\section*{A Survey of Young MC Remnants}

In this section I survey the properties and nature of the known young
supernova remnants in the MCs.  Objects are introduced in
chronological order according to when each was identified as a young
SNR, based on the criteria just described.

\underbar{N132D}

One of the original Large Magellanic Cloud (LMC) SNRs, N132D, can be
considered to be young.  Originally noted as an optical emission
nebula \cite{henize} and a continuum radio source
\cite{mathewsonhealy64}, N132D was confirmed as a remnant through
optical spectroscopy \cite{westerlundmathewson66}.  In 1976
oxygen-rich optical filaments spanning a wide velocity range
($\sim$4000 km s$^{-1}$) were discovered in the center of N132D
\cite{danzigerdennefeld76}. Although this is unquestionably emission
from SN ejecta, N132D also shows an outer shell of normal swept-up
interstellar medium (ISM) that in fact dominates the integrated X-ray
emission \cite{hughes87} and appears to be associated with a giant
molecular cloud \cite{banas97}.  N132D was almost surely a
core-collapse SN, but no compact remnant has yet been detected.  The
age of the remnant estimated from the expansion and size of the
oxygen-rich filaments is $\sim$3000 yrs, while an X-ray spectral
analysis yields 2000--6000 yrs \cite{hhk98}.  Preliminary results from
\chandra\ HETG \cite{canizares01} and \xmm\ \cite{behar01} observations
of N132D are discussed elsewhere in this volume.  The EPIC MOS
narrowband images of N132D show strong oxygen emission in the center,
while at higher energies the most intense emission comes from the
southern limb where the remnant is interacting with dense material
presumably associated with the molecular cloud that lies there.

\underbar{0540$-$69.3}

It was not until the launch of the \EO\ that additional examples of
young remnants were discovered in the MCs. Three of the four new young
remnants were discovered through their spatially-extended, soft X-ray
emission and later identified as optical and radio emission nebulae.
(More on them below.)
The fourth remnant, SNR 0540$-$69.3 (also
referred to as N158A), was known to be a nonthermal radio source
\cite{lemarne} and had been suggested to be a SNR as early as 1973
\cite{mathewsonclarke73}.  High velocity oxygen-rich filaments with
velocities spanning $\sim$3000 km s$^{-1}$ were discovered
\cite{mathewson80} soon after the \einstein\ survey of the LMC showed
SNR 0540$-$69.3 to be the third brightest extended soft X-ray source
in the Cloud \cite{longhelfand79}. The X-ray emission is almost entirely
nonthermal \cite{clark82}, like the Crab Nebula, and indeed SNR
0540$-$69.3 also harbors a rapidly spinning compact remnant, in this
case a 50-ms pulsar \cite{seward84}.  The spin-down age of the pulsar
$\tau = P/2\dot P$ is roughly 1600 yr, while the kinematic age is
$\sim$800 yr \cite{kirshner89}.  This remnant is probably the second
youngest SN in the LMC, only exceeded in its youth by SN1987A.

Within the last year \chandra\ HRC observations \cite{gotthelf00} have
revealed the fine-scale structure of the remnant: a patchy incomplete
shell surrounding an elliptically-shaped plerionic core that greatly
resembles the Crab Nebula (Fig.~1). In the optical band the core, as
imaged by \hst, shows a beautiful filamentary structure with complex
spectral variations with position \cite{morse01}.  The ACIS-S CCD
spectral data (Fig.~1) show that the shell is swept-up ISM, as opposed
to SN ejecta.  A nonequilibrium ionization thermal model yields
abundances for O, Ne, Mg, Si, and Fe of $\sim$0.3 times solar,
consistent with the low metallicity of the LMC.  Under the plausible
assumption that this emission represents the blast wave, then our
fitted electron temperature of $\sim$0.4 keV is remarkably low for
such a young remnant. The X-ray spectrum also requires a hard power
law tail with a photon index of $\alpha \sim 2.5$.  Whether this
emission is related to the pulsar itself or is an indication of
nonthermal emission at the shock front, like in SN1006 and a few other
Galactic remnants, remains to be determined. In any event, as the
unique example of an oxygen-rich remnant that also contains a young
pulsar, SNR 0540$-$69.3 is an important object for studying the link
between pulsars and the massive stars that form them in core-collapse
SNe.

\begin{figure} 
\centerline{
\epsfig{file=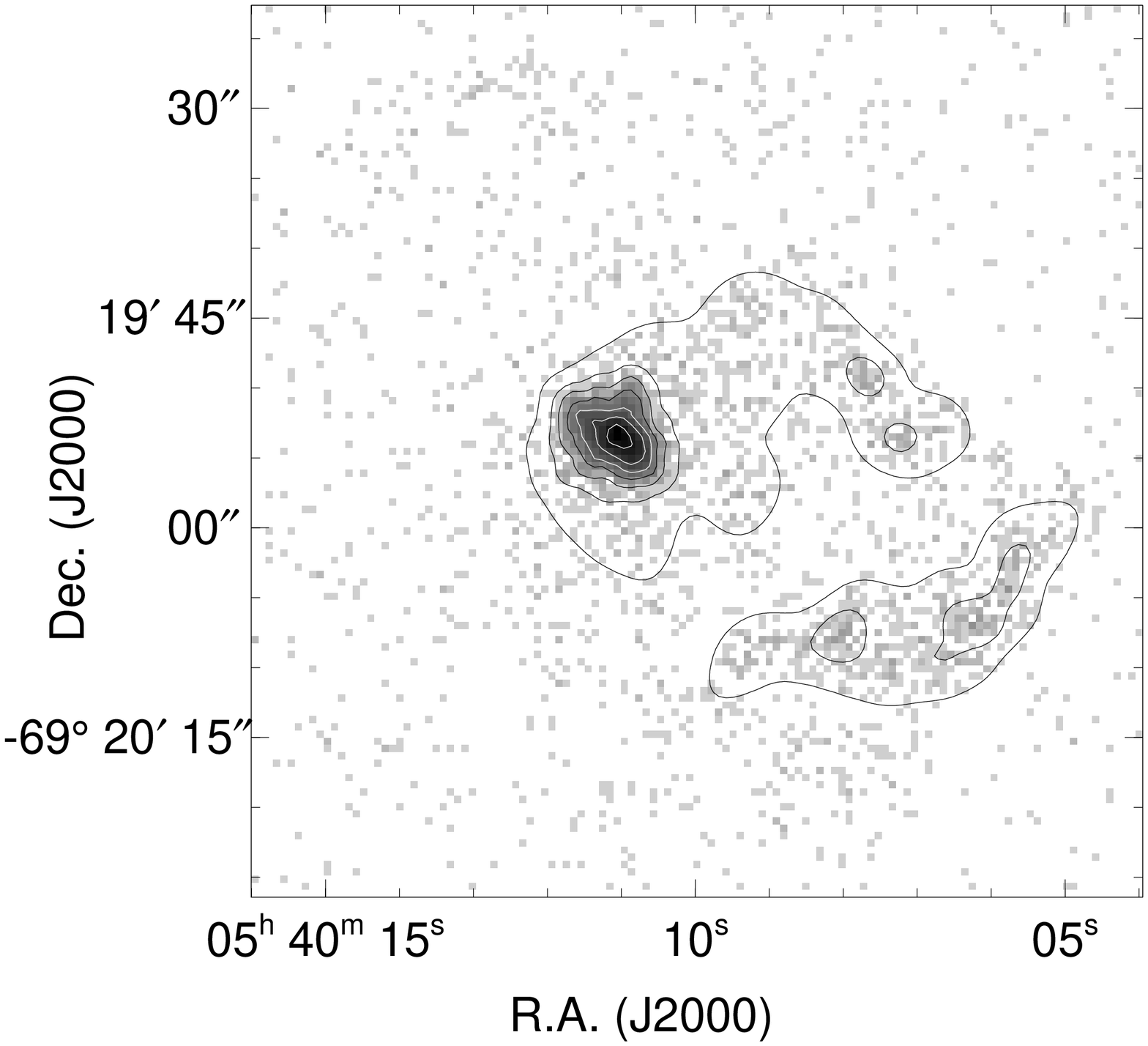,height=2.2in,width=2.2in}
\epsfig{file=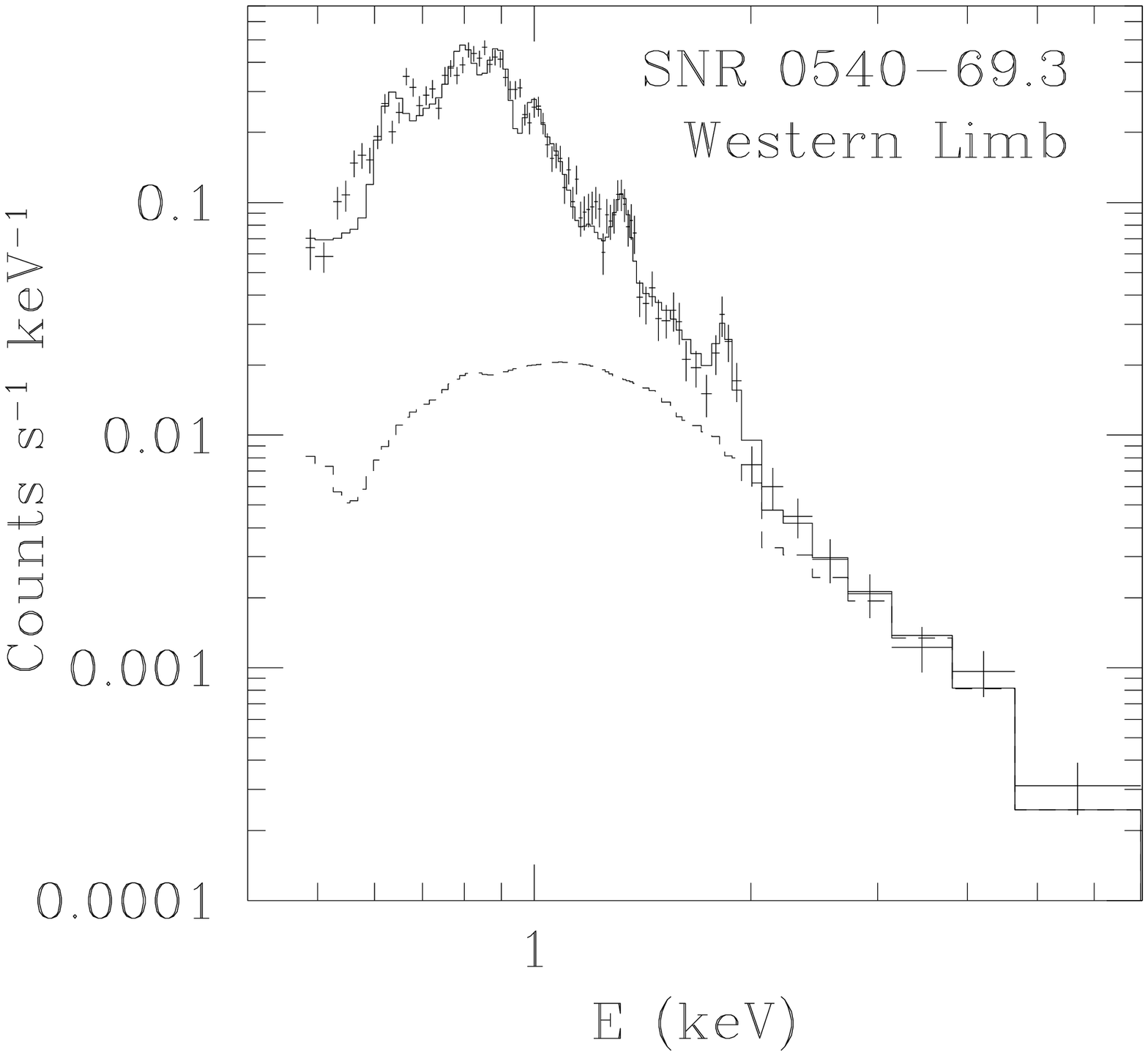,height=2.05in,width=2.05in}
}
\vspace{10pt}
\caption{Left panel shows the HRC image of SNR 0540$-$69.3 as both raw counts
per pixel (grayscale) and adaptively smoothed contours.  Right panel
shows the ACIS-S CCD spectrum of the faint western limb with thermal model
and power law (dashed). Elemental abundances are consistent with swept-up ISM
rather than SN ejecta.}
\label{fig1}
\end{figure}


\underbar{1E0102.2$-$7219}

Yet another oxygen-rich remnant was the next young one to be
identified in the Clouds.  This was also the result of an \einstein\
X-ray survey, this time of the Small Magellanic Cloud (SMC)
\cite{seward81}.  The new remnant, 1E0102.2$-$7219, is the second
brightest soft X-ray source in the SMC. Optical follow-up confirmed
the identification as a SNR and revealed it to be extremely
oxygen-rich \cite{dopita81} with [O {\sc iii}] emission covering
$\sim$6500 km s$^{-1}$ in velocity \cite{tuohy83}.  From this a
kinematic age of $\sim$1000 yr was estimated.  \asca\ showed the X-ray
emission from 1E0102.2$-$7219 to be dominated by lines from He- and
H-like species of O, Ne, Mg, and Si, although it was not possible to
obtain a simple interpretation of the integrated spectrum
\cite{hayashi94}.  No compact remnant, either pulsar or plerionic
core, has been detected down to a flux limit of $10^{34} \rm\, erg
\,cm^{-2}\, s^{-1}$ \cite{gaetz00}. However, the ACIS-S CCD data did
reveal a faint shelf of X-ray emission presumably from the blast wave
lying beyond the bright rim of oxygen-rich ejecta \cite{gaetz00}.  As
for the ejecta, \chandra\ HETG line images from the past year reveal
high velocity O- and Ne-rich X-ray emitting ejecta with different
positions in the shell displaying different red- and blue-shifts
\cite{canizares01}.  The remnant's size depends on the emission line
under study.  In general the remnant appears larger in higher
ionization species than it does in lower ones, an effect that is also
seen in the lower spectral resolution ACIS-S CCD data \cite{gaetz00}.
One simple interpretation is that we are seeing the progressive
increase in ionization caused by the reverse shock propagating inward
through the ejecta.  The outermost ejecta material was heated by the
reverse shock first and hence has had the most time to ionize up to
higher ionization species.  The most recently shocked ejecta, however,
lies interior to this and should show a lower ionization state.
Although qualitatively consistent with the observations, this picture
may be oversimplified.  In addition to the temporal effect just
described, the temperature and density in the ejecta may vary
significantly from the contact discontinuity (outside) to the reverse
shock (inside) and thereby cause spectral variations with
position. The temperature and density variations contain information
about the original distribution of matter in the ejecta and the
circumstellar medium (CSM) (see, e.g., \cite{chevalier82}). Hopefully,
further study will allow us to disentangle the effects of time,
density, and temperature and learn more about the ejecta and CSM in
1E0102.2$-$7219.

Very recently my collaborators and I used the \chandra\ ACIS-S CCD
data on 1E0102.2$-$7219 to address some basic issues of shock physics
\cite{hughes00}.  The question we originally set out to investigate
was ``To what extent are electrons heated at high Mach number shocks
in SNRs?'' Do the electrons rapidly attain similar temperatures to the
ions through anomalous heating processes driven by plasma
instabilities at the shock front (see, for example, \cite{cargill88})?
Or do the ion and electron temperatures initially differ by their mass
ratio ($m_p/m_e = 1836$) with the electrons subsequently gaining heat
slowly from the ions through Coulomb collisions?

Clearly the information needed for this study was the post-shock
temperatures of both the electrons and ions. We determined the
post-shock electron temperature from fits to the X-ray spectrum of the
blast wave region that \chandra\ had revealed for the first time.
Fig.~2 shows this spectrum and for comparison the spectrum of a
portion of the bright rim of ejecta which displays a markedly
different spectral character.  The derived abundances from the
blast-wave spectrum confirm its origin as swept-up ISM and, for a
variety of nonequilibrium ionization spectral models, the derived
electron temperature was constrained to be less than 1 keV.  It is
impossible at the present time to determine the post-shock ion
temperature directly (since there is no H line emission from the
remnant). We were, however, able to determine the velocity of the
blast wave by measuring the angular expansion rate of the
remnant. Previous images of 1E0102.2$-$7219 made by \rosat\ and
\einstein\ were compared to the \chandra\ image and it was found that
a significant change in size had occurred (Fig.~2).  From the
expansion rate (0.100\% $\pm$ 0.025\% yr$^{-1}$) and known distance a
blast wave velocity of $\sim$6000 km s$^{-1}$ was determined.  If we
assume that the electron and ion temperatures are fully equilibrated
at the shock front, then the post-shock electron temperature we
measure is about a factor of 25 lower than expected given this blast
wave velocity.  How low can the post-shock electron temperature be if
the post-shock ion temperature is given by the Rankine-Hugoniot jump
conditions?  This is set by assuming that post-shock electrons and
ions exchange energy solely through Coulomb interactions. In this
case, we find that the minimum expected electron temperature is
$\sim$2.5 keV, still significantly higher than what we measure.  Our
explanation for this discrepancy is that efficient shock acceleration
of cosmic rays has reduced the post-shock temperature of both the
electrons and ions \cite{blanford87}, or in other words, a large
fraction of the shock energy has been diverted from the thermal
particles and has instead gone into generating relativistic particles,
which produce the remnant's radio emission \cite{amyball93}.
Nonlinear models for efficient shock acceleration of cosmic rays
\cite{ellison00} predict mean post-shock electron and ion temperatures
of 1 keV for high Mach number shocks, i.e., 100--300, as appropriate
to 1E0102.2$-$7219, lending strong quantitative support for this
scenario.


\begin{figure} 
\centerline{
\epsfig{file=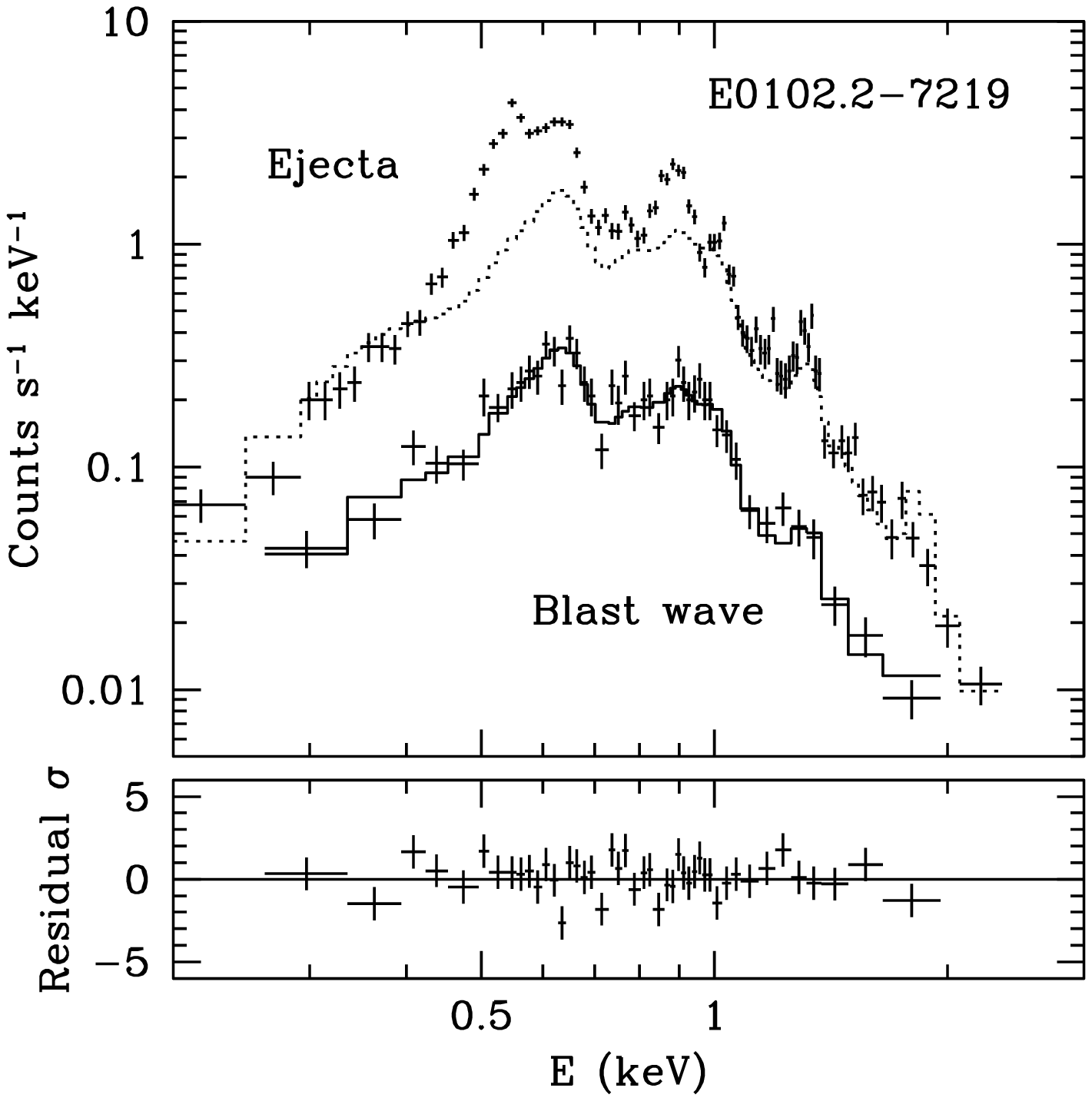,height=2.2in,width=2.2in}
\epsfig{file=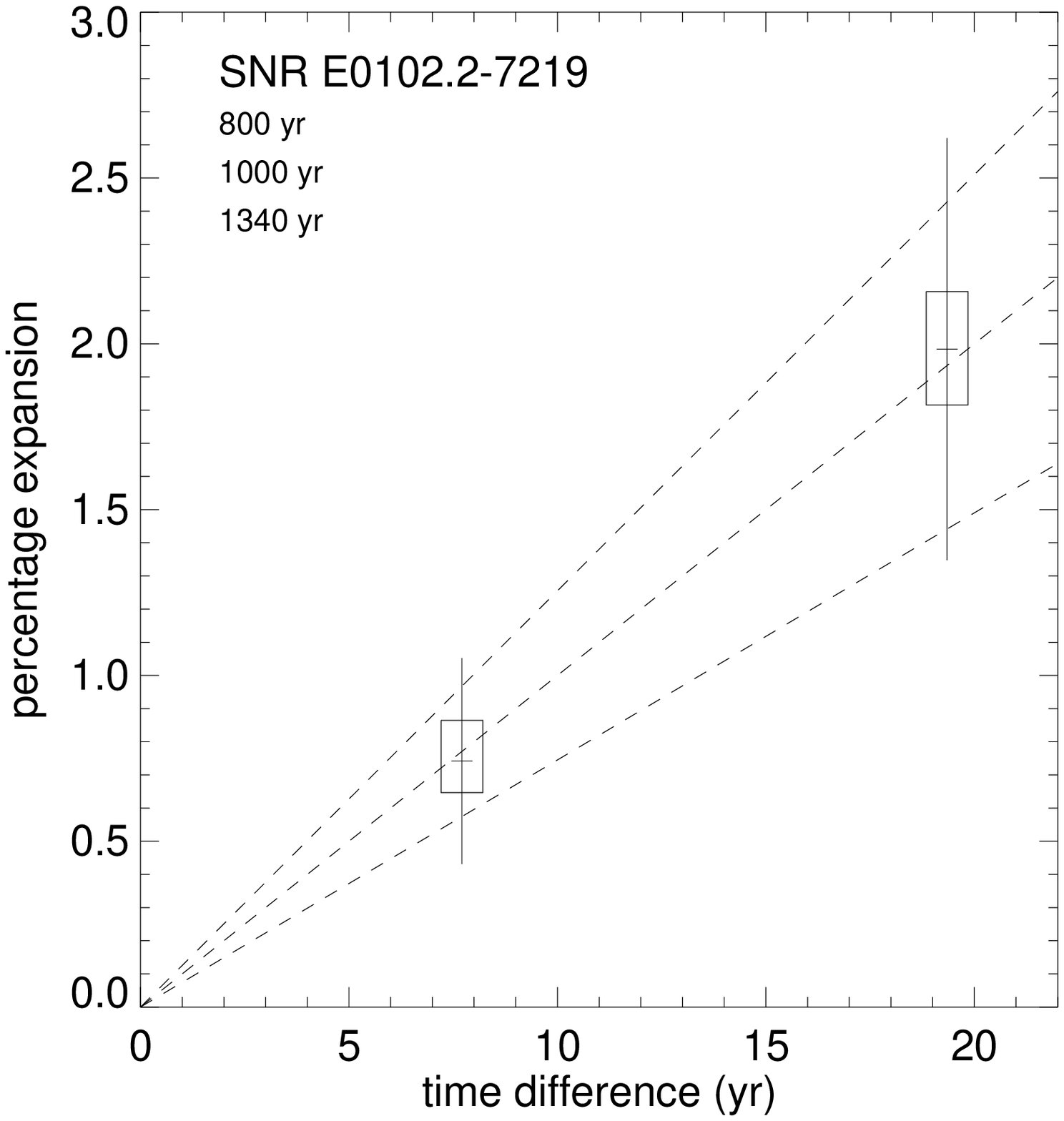,height=2.3in,width=2.3in}
}
\vspace{10pt}
\caption{Left panel shows the ACIS-S CCD spectra of a portion of the
outer blast wave and bright rim of ejecta in SNR 1E0102.2$-$7219.
Right panel shows the expansion rate of this remnant from a comparison
of the \chandra\ ACIS image with earlier \rosat\ and \einstein\
images.}
\label{fig2}
\end{figure}

\underbar{E0509$-$67.5 and E0519$-$69.0}

E0509$-$67.5 and E0519$-$69.0 are two apparently similar SNRs in the
LMC that were discovered through their X-ray emission \cite{lhg81}.
Their X-ray morphologies are strongly shell-like and reasonably
symmetric.  Optical follow-up \cite{tuohy82} revealed them to be
members of a rare class of faint SNRs, like Tycho and SN1006 in the
Galaxy, that are dominated by Balmer line emission and show little or
no forbidden line emission from, e.g., [O {\sc iii}] or [S {\sc ii}].
Like their Galactic cousins, these two LMC SNRs were originally
believed to be young in part because they are among the smallest
remnants in the LMC with diameters of 7 pc (E0509$-$67.5) and 8 pc
(E0519$-$69.0). In addition for E0519$-$69.0, at least, there is
evidence for a broad component to H$\alpha$ with FWHM velocity widths
ranging from 1300 km s$^{-1}$ \cite{smith91} to 2800 km s$^{-1}$
\cite{tuohy82}.  This emission arises from charge exchange between
fast protons and neutral hydrogen atoms in the immediate post-shock
region \cite{chevalier78} with the width of the line being related to
the shock velocity and assumptions about the post-shock partition of
energy into hot electron, ions, and cosmic rays \cite{chevalier80}.
The broad line in E0519$-$69.0 implies a probable shock velocity range
of 1000--1900 km s$^{-1}$ (although there are likely to be significant
variations with position) and an age of 500--1500 yrs
\cite{smith91}. No broad H$\alpha$ has yet been detected from
E0509$-$67.5 which has been interpreted as implying a lower limit on
the actual shock velocity in the remnant of $>$2000 km s$^{-1}$ with a
corresponding upper limit on the age of $<$1000 yr \cite{smith91}.
\asca\ observations of these remnants \cite{hughes95} revealed the
spectacular ejecta-dominated nature of their integrated X-ray
emission.  The spectra are dominated by a broad quasi-continuum
emission of unresolved Fe L-shell lines around 0.7--0.9 keV as well as
strong K$\alpha$ lines of highly ionized Si, S, Ar, and Ca.  The
spectra were shown to be qualitatively consistent with the
nucleosynthetic products expected from a Type Ia SN explosion, thereby
significantly strengthening the connection between Balmer-dominated
SNRs and the remnants of Type Ia SN.

Early \chandra\ results on E0519$-$69.0 \cite{williams01} show an
integrated spectrum quite similar to the \asca\ one, but the lower
effective background of the \chandra\ data allows for clear detection
of the important Fe K$\alpha$ line.  There are spatial/spectral
variations across the remnant with the most notable difference being a
strong increase in emission in the 0.5--0.7 keV band at the rim.

The \chandra\ broad-band X-ray image of E0509$-$67.5 is shown in
Fig.~3 along with the integrated spectrum. The shell is quite round,
nearly complete, and brightest on the western side; all of which are
quite similar to the H$\alpha$ morphology \cite{smith91}.  The
intensity fluctuations around the bright limb of the remnant are
generally significant with peak to valley brightness variations of
about a factor of two on spatial scales of just a few arcseconds.
Whether this apparent clumping of the ejecta originated during the SN
explosion process itself, arose later on through Rayleigh-Taylor
instabilities from interaction with the ambient medium, or has some
other origin remains to be investigated. There is no morphological
evidence for a double shock structure: the X-ray emission is dominated
by metal-rich ejecta and no outer blast wave component is detected.
Again the spectrum is quite similar to the \asca\ one \cite{hughes95},
although here with \chandra\ we have detected Ar, Ca, and Fe K$\alpha$
line emission at energies of 3.0 keV, 3.7 keV, and 6.4 keV,
respectively.  The Si and S K$\alpha$ lines in the integrated spectrum
of E0509$-$67.5 are quite strong (equivalent widths of 1.9 keV for Si
and 1.2 keV for S), very broad (FWHM line widths of 100 eV and 120
eV), and appear at much lower energies than the He- and H-like
K$\alpha$ lines should (mean line energies are 1.804 keV and 2.377
keV).  There is no change in the width or mean energy of the Si
K$\alpha$ line with radius, which argues against velocity effects
(which would have to be enormous anyway) as the source of the line
broadening. The most likely explanation for the broad Si line is that
it is a blend of lines from a wide range of charge states, i.e., from
Si$^{12+}$ (He-like) to Si$^{5+}$ or lower. A similarly broad range of
charge states for S is required to explain that broad line.  It seems
likely that a range of temperatures or ionization timescales will be
necessary to account for the broad lines; simple planar shock models
tend to produce too narrow a distribution of charge states.  The
spectral analysis with radius does reveal one significant
variation---the Si line equivalent width drops by a factor of 2 (to a
value of $\sim$1 keV) right at the edge of the remnant---perhaps an
indication of some dilution of the ejecta emission by a component of
swept-up ISM.

\begin{figure} 
\centerline{
\epsfig{file=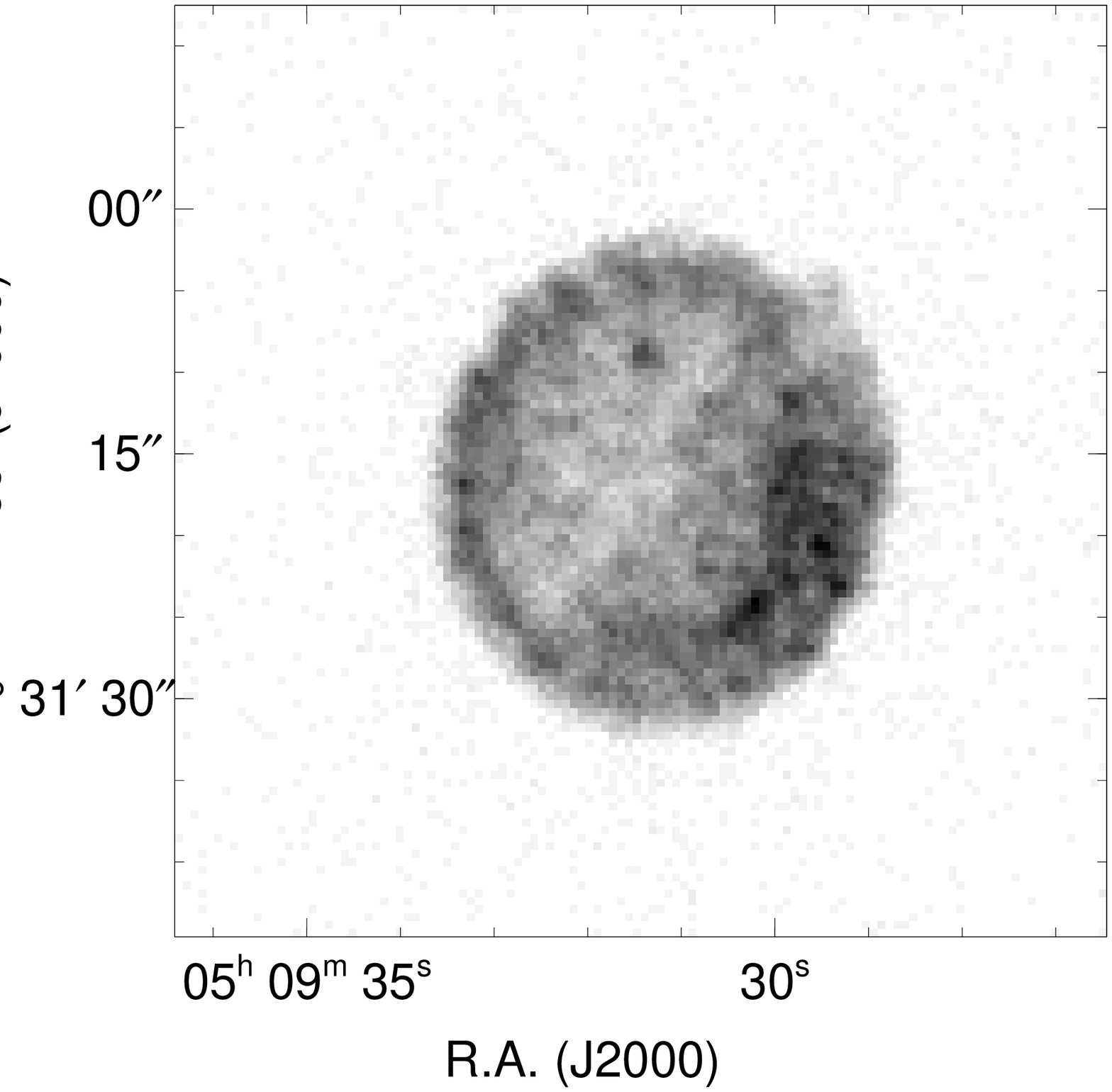,height=2.35in,width=2.35in}
\epsfig{file=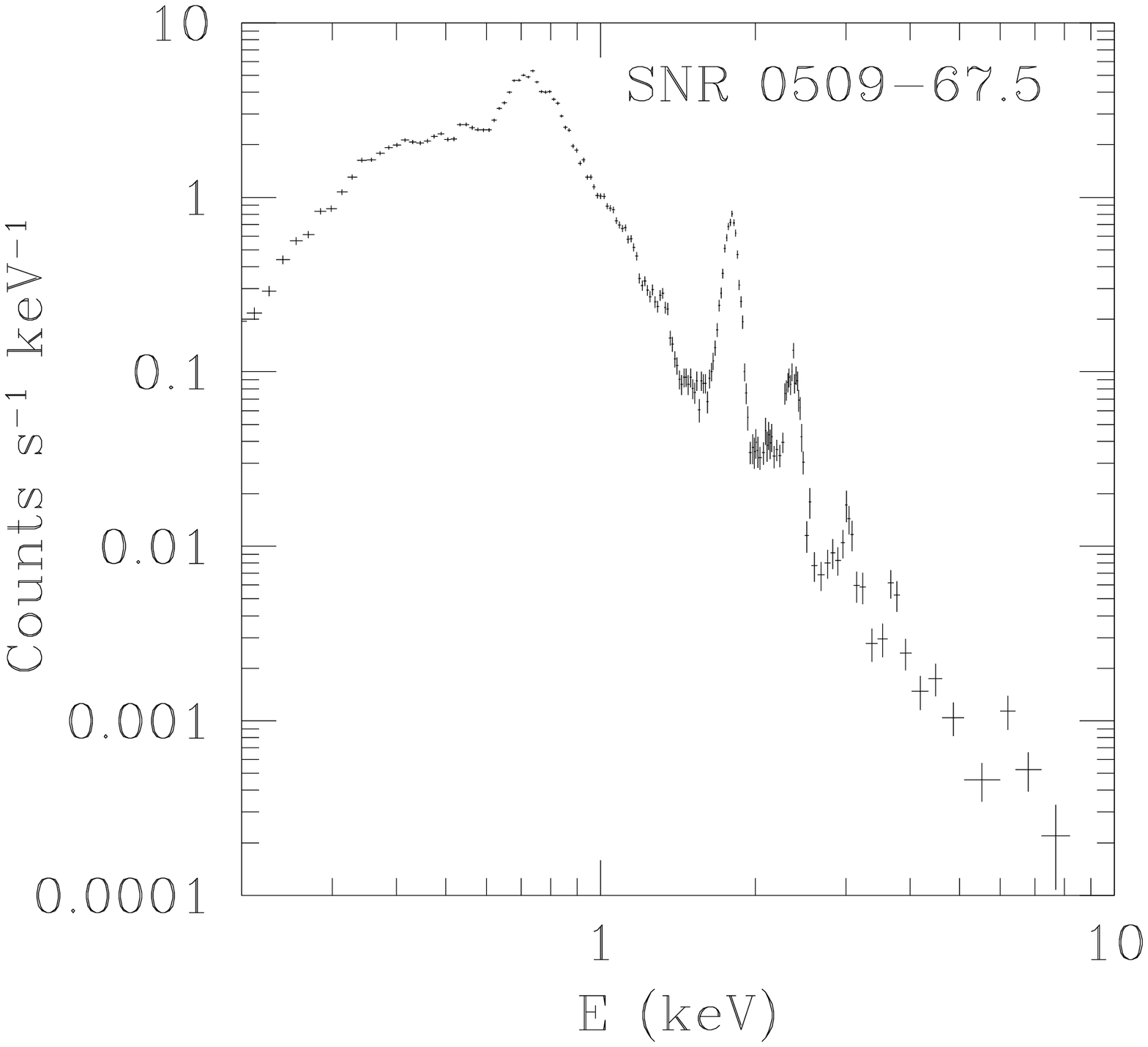,height=2.26in,width=2.26in}
}
\vspace{10pt}
\caption{Left panel shows the ACIS-S CCD image of E0509$-$67.5 over the
0.2--7.0 keV band. Right panel shows the integrated spectrum of the
remnant.}
\label{fig3}
\end{figure}

\underbar{N103B}

First identified in 1973 as a SNR based on its radio and optical
properties \cite{mathewsonclarke73}, N103B was detected as an
extended, soft X-ray source by the \EO\ \cite{longhelfand79}.  It is
the fourth brightest X-ray remnant in the LMC and is also one of the
smallest LMC SNRs ($\sim$7 pc in diameter).  Optically the remnant
consists of several small, bright knots visible in H$\alpha$, [O {\sc
iii}], and [S {\sc ii}].  Its proximity to the star cluster NGC 1850
and the H {\sc ii} region DEM 84 led to the suggestion that SNR N103B
had a population I progenitor \cite{chu88}.  It was not until \asca\
that N103B was revealed to be young based on its ejecta-dominated
spectrum \cite{hughes95}. The surprising result was that the X-ray
spectrum of N103B bears a remarkable similarity to the
Balmer-dominated SNR E0519$-$69.0 and indeed both remnants are now
believed to be remnants of Type Ia SN.  The \chandra\ data
\cite{lewis01} show a highly structured remnant that is much brighter
on the western limb than toward the east.  On large scales it
resembles the radio remnant \cite{dickel95} but on finer spatial
scales the bright radio and X-ray emission features tend not to
overlap. There are complex spatial/spectral variations across the
remnant including changes in the equivalent width of the He-like Si
K$\alpha$ emission line which appears to increase radially outward.
In contrast to E0509$-$67.5, the Si K$\alpha$ line from N103B shows
both a He- and H-like component indicating that N103B is at a higher
mean ionization state.

\underbar{N157B}

Originally detected as a nonthermal radio source \cite{lemarne}, and
suggested as a possible SNR in 1973 \cite{mathewsonclarke73}, it was
not until extended soft X-rays were detected by the \EO\
\cite{longhelfand79} that N157B (also referred to as 30 Dor B) was
confidently confirmed as a remnant. Its featureless powerlaw X-ray
spectrum \cite{clark82} and flat radio spectral index ($\alpha \sim
-0.19$ \cite{mathewson83} \cite{lazendic00}) were taken as strong
evidence that the remnant was Crab-like and should contain a rapidly
rotating pulsar. The pulsar in N157B was discovered in 1998
\cite{marshall98} with a spin period of 16.1 ms making it the fastest
known pulsar in a SNR.  The spin-down rate of $\dot P = 5.126\times
10^{-14} \,\rm s\, s^{-1}$ implies a characteristic age of $\tau
=5000$ yr which agrees well with other estimates of the remnant's age
\cite{wang98}.

\underbar{DEM71}

A new example of a possible young SNR is DEM71. Initially noted as an
optical nebulosity \cite{dem76}, it was detected as an extended soft
X-ray source by \einstein\ and proposed as a SNR \cite{lhg81}.
Optical follow-up revealed it to be a Balmer-dominated remnant
\cite{tuohy82} with broad H$\alpha$ line emission
\cite{smith91}. Given its size (20 pc diameter) and shock velocity of
$\sim$500 km s$^{-1}$ (estimated from the broad H$\alpha$ line), the
remnant is $\sim$8000 yr old.  \asca\ spectroscopy revealed the first
tantalizing hints of youth for this SNR in the form of an enhanced
abundance of Fe \cite{hhk98}.

\begin{figure} 
\centerline{
\epsfig{file=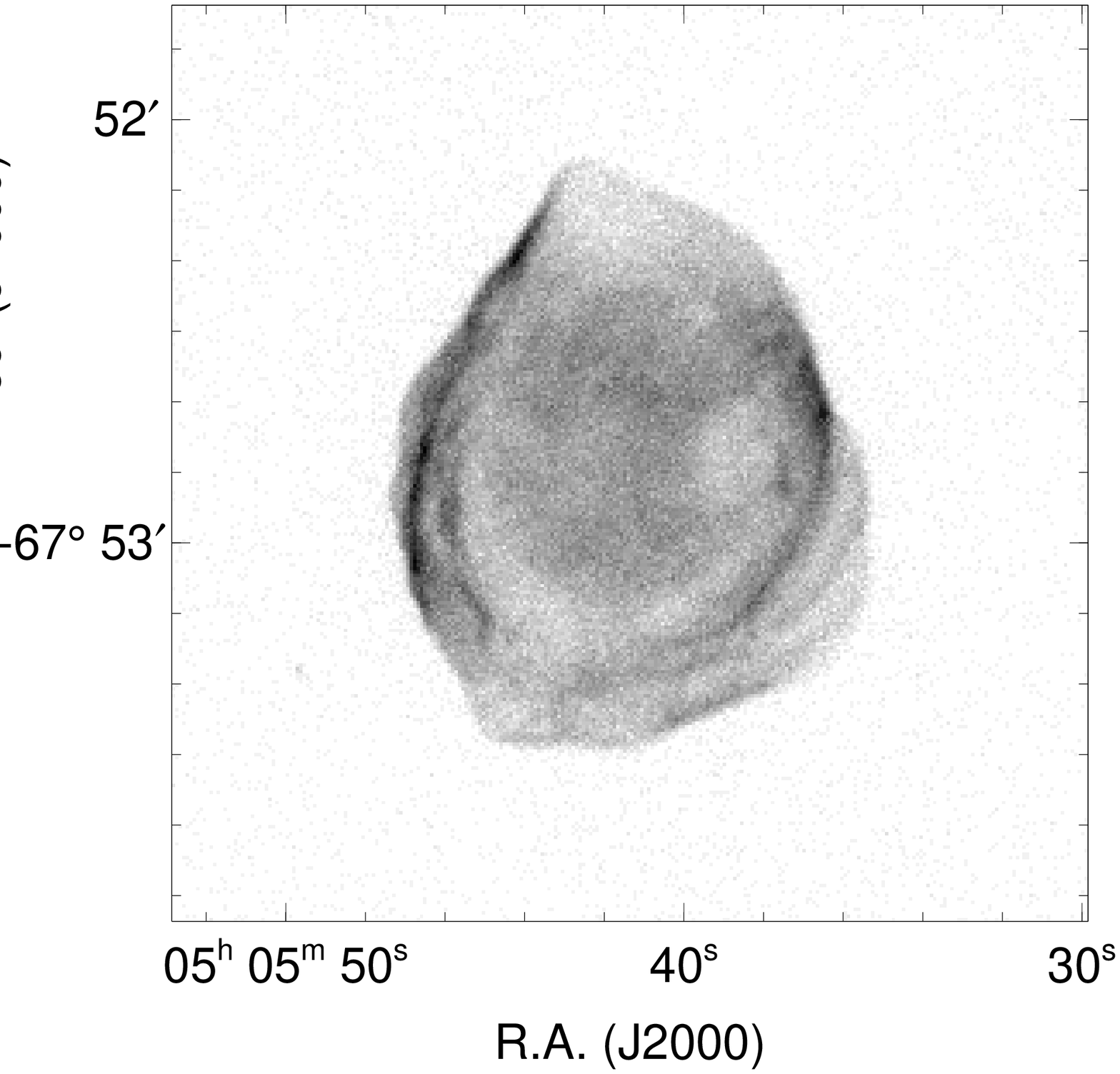,height=2.35in,width=2.35in}
\epsfig{file=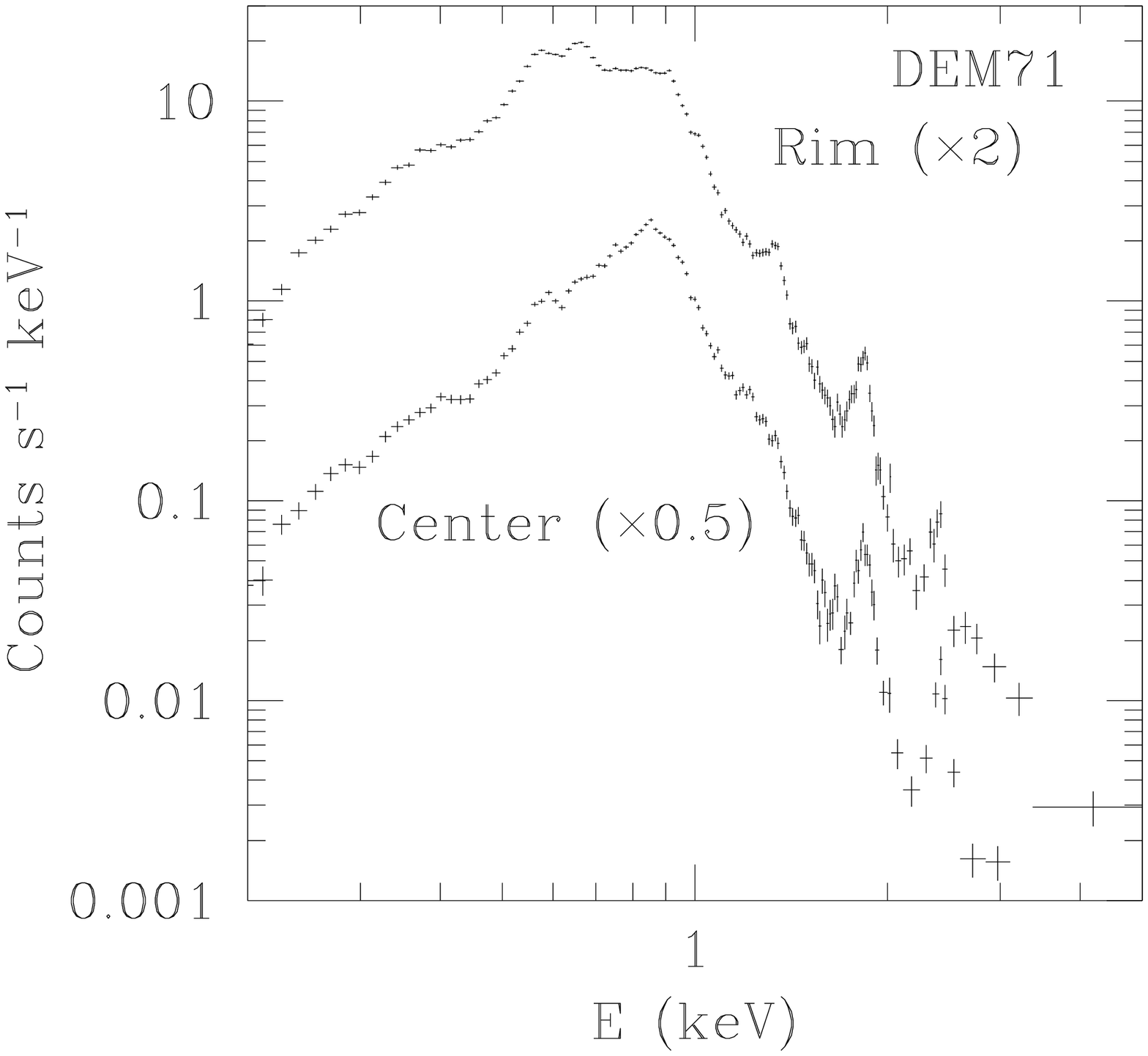,height=2.2in,width=2.2in}
}
\vspace{10pt}
\caption{Left panel shows the ACIS-S CCD image of DEM71 over the
0.2--4.4 keV band. Right panel shows the spectrum of the
remnant from the outer rim and the central region.}
\label{fig4}
\end{figure}

The broadband image of DEM71 from \chandra\ (fig.~4) shows emission
from an outer rim that matches nearly perfectly the optical H$\alpha$
image, including the brightness of the eastern and western rims as
well as the multiple filaments along the southern edge.  The nearly
circular, faint, diffuse emission that almost fills the interior has
no counterpart in the optical band \cite{mathewson83}.  The ACIS-S CCD
spectrum of the center (Fig.~4) is dominated by a broad
quasi-continuum of emission near 1 keV from Fe L-shell lines, plus Si
and S K$\alpha$ lines at higher energy.  This spectrum is grossly
different from that of the outer rim, which is generally softer and
shows prominent O and Mg K$\alpha$ lines.  We propose that the rim
represents the blast wave moving out into the ambient interstellar
medium, while the interior emission is the reverse shock propagating
through iron-rich ejecta. Detailed study of the \chandra\ data is now
underway using the outer rim emission to study the basic physics of SN
shocks and the central Fe-rich emission to investigate the evolution
of metal-rich ejecta in SNRs.

\section*{Conclusions}

%

I expect the list of ``young'' remnants in the MCs to grow as
improvements in sensitivity make it easier to identify youthful
aspects of older known remnants.  It is likely that there are still a
few pulsars left to be discovered in the MC SNRs either through direct
measurement of pulsed radiation or from the identification of their
associated pulsar-wind nebulae. If DEM71 can be considered to be a
guide, then even remnants that are several thousand years old can show
evidence for SN ejecta. N63A, N49, N23 in the LMC, all of which are
smaller in size than N132D and DEM71, are in this age range and
therefore have the potential for showing traits of youth.  It will be
worth keeping an eye on these remnants in the coming years.

%
%
%
%
%

{\it Acknowledgments} I gratefully acknowledge Dave Burrows, Anne
Decourchelle, Gordon Garmire, Parviz Ghavamian, Karen Lewis, John
Nousek, Cara Rakowski, and Pat Slane for their collaborations on
various studies of MC SNRs.  I would also like to thank Andy Rasmussan
and Ehud Behar for sharing XMM results prior to publication. This work
was partially supported by \chandra\ General Observer grant GO0-1035X.

\end{document}